\newcommand{\CLASSINPUTtoptextmargin}{1in}
\newcommand{\CLASSINPUTbottomtextmargin}{1in}
\documentclass[conference, onecolumn]{IEEEtran}

\usepackage[latin9]{inputenc}
\usepackage{textcomp}
\usepackage{amsmath}
\usepackage{amssymb}
\usepackage{graphicx}
\usepackage{tabularx,ragged2e,booktabs,caption}
\usepackage{epsfig,rotating,setspace,latexsym,amsmath,epsf,amssymb,bm}
\usepackage{cite,graphicx,color,subfigure, amsthm}
\makeatletter
\usepackage[english]{babel}

\makeatother

\usepackage{babel}
\ifCLASSINFOpdf
\else
\fi

\usepackage{balance}

\IEEEoverridecommandlockouts
\begin{document}
\title{Ultra-Reliable Cloud Mobile Computing with Service Composition and Superposition Coding}

\author{\IEEEauthorblockN{Seyyed Mohammadreza Azimi and Osvaldo Simeone \thanks{The work of S.M. Azimi and O. Simeone was partially supported by the U.S. NSF through grant 1525629.}}
\IEEEauthorblockA{CWCSPR, ECE Dept.\\New Jersey Institute of Technology\\Newark, NJ\\
Email: \{sa677, osvaldo.simeone\}@njit.edu}
\and
\IEEEauthorblockN{Onur Sahin}
\IEEEauthorblockA{InterDigital\\London, UK\\Email: onur.sahin@interdigital.com}
\and
\IEEEauthorblockN{Petar Popovski\thanks{The work of P. Popovski has been in part supported by the European Research Council (ERC Consolidator Grant Nr. 648382 WILLOW) within the
Horizon 2020 Program.}}
\IEEEauthorblockA{Aalborg University\\
Aalborg, Denmark\\Email: petarp@es.aau.dk}}


%Telephone: (800) 555--1212, Fax: (888) 555--1212}


\maketitle

\begin{abstract}
An emerging requirement for 5G systems is the ability to provide wireless ultra-reliable communication (URC) services with close-to-full availability for cloud-based applications. Among such applications, a prominent role is expected to be played by mobile cloud computing (MCC), that is, by the offloading of computationally intensive tasks from mobile devices to the cloud. MCC allows battery-limited devices to run sophisticated applications, such as for gaming or for the ``tactile'' internet. This paper proposes to apply the framework of reliable service composition to the problem of optimal task offloading in MCC over fading channels, with the aim of providing layered, or composable, services at differentiated reliability levels. Inter-layer optimization problems, encompassing offloading decisions and communication resources, are formulated and addressed by means of successive convex approximation methods. The numerical results demonstrate the energy savings that can be obtained by a joint allocation of computing and communication resources, as well as the advantages of layered coding at the physical layer and the impact of channel conditions on the offloading decisions.
\end{abstract}


\IEEEpeerreviewmaketitle

\section{Introduction}
An emerging requirement for 5G systems is the ability to provide wireless ultra-reliable communication (URC) services with close-to-full availability for cloud-based applications (see, e.g., \cite{METIS}). Among such applications, a prominent role is expected to be played by mobile cloud computing (MCC), that is, by the offloading of computionally intensive tasks from mobile devices to the cloud \cite{MCC}. MCC allows battery-limited devices to run sophisticated applications, such as for video processing, object recognition, gaming, automatic translation and medical monitoring, and can be an enabler of the ``tactile'' internet \cite{Fett,Dohler}. Well-known applications that are based on MCC include Google Voice Search and Apple Siri.

\begin{figure}[t]
\centering \includegraphics[width=0.4\textwidth]{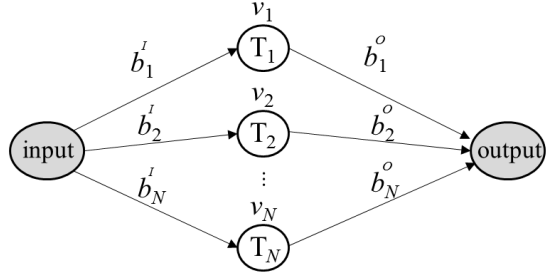} \protect\caption{ An example of a call
graph in the class of map-reduce applications under study. Gray nodes correspond to tasks that must be run at the mobile.}\label{Fig1}

\label{fig:system}

\end{figure}

Existing solutions for the optimization of the offloading decisions for MCC generally abstract the contribution of the underlying communication network by assuming reliable links with \emph{fixed} achievable rates (see, e.g., \cite{MCC,Hermes} and references therein). More recently, it was recognized that there is an important interplay between the offloading decisions at the application layer and the operation of the underlying communication network, which can provide different trade-offs between rate and energy expenditure at the mobile devices. As a result, the inter-layer optimization of offloading decisions and communication network parameters, such as transmission powers, were studied in \cite{Barbarossa,Scutari1} and references therein, as well as in \cite{Barbarossa1,Khalili}. In this line of work, the focus is on the resource allocation of communication and computing functionalities, and a key assumption is the reliability of the communication links at the rates specified by current channel conditions and by the power allocation. Furthermore, the applications to be offloaded can be assumed to be unsplittable as in \cite{Barbarossa} or splittable into constituent subtasks as in \cite{Barbarossa1,Khalili}.

The assumption of reliable communication is in practice too strong when communication takes place over wireless fading channels, especially when latency constraints prevent the use of retransmission protocols to reduce the probability of error. In light of this important motivation, this work aims at studying the problem of joint optimization of offloading decisions and communication system's parameters by accounting for the limited reliability of fading channels with given diversity degrees.

At the application layer, we postulate, as in \cite{Popovski} (see also \cite{Dohler}), that certain applications can be designed so as to ensure \emph{service composition}: the application can be run at different levels of accuracy or quality of experience, with higher levels requiring a larger number of CPU cycles. For example, in an object recognition application based on video or image frames, the first service level may correspond to identification of dangerous obstructions, the second to the recognition of landmarks, the third to the search of businesses of possible interest, etc. We observe that the idea of service composition is already implemented in scalable video coding and, more generally, in successive refinement data compression. When coupled with transmission with differentiated reliability levels on the communication network, the approach will be referred to as \emph{reliable service composition} \cite{Popovski}.

This paper proposes to apply the framework of reliable service composition to the problem of optimal task offloading in MCC over fading channels, with the aim of providing layered, or composable, services at differentiated reliability levels. We focus on a simple application call graph, exemplified in Fig. \ref{Fig1}, which is related to the popular ``map-reduce'' programming model, in which multiple parallel tasks operate between an input task that prepares the input (``map") and an output task that combines the outputs of the parallel tasks ``reduce"). In MCC, each one of the parallel task may be offloaded or not. The application is designed, according to the service composition principle, so that running the first task $\mathrm{T}_1$, along with input and output tasks, ensures the basic level of service, while the execution of successively more tasks $\mathrm{T}_2,\mathrm{T}_3,...$ allows a higher-accuracy outcome to be obtained. As an example, the parallel tasks may correspond to the processing of different features to extract sufficient statistics in a detection application.

For communication, we consider and compare both time division (TD) transmission and superposition coding (SC), where the latter has been widely studied for the transmission successive refinement compression layers \cite{Gunduz}. Inter-layer optimization problems, encompassing offloading decisions and communication resources, are formulated and addressed by means of successive convex approximation methods \cite{Scutari}.

 The paper is organized as follows. In Section II, first the system model is described and then optimization
algorithms over both offloading decisions and communication parameters for TD and SC transmission are provided. Numerical results are provided in Section III and the paper is concluded in Section IV.

\section{System Model and Problem Formulation}
In this section, we define system model and problem formulation.

\subsection{System Model}
We focus on the optimization of offloading decisions and communication parameters for a given mobile user, which can communicate to a base station (BS) via a fading wireless channel. The BS is in turn connected to a cloud processor. As discussed, the application to be run at the mobile is characterized by a set of processing tasks that could be run locally or remotely at the cloud.

\textbf{Call Graph:} A call graph is used to describe the relationship between computing tasks (e.g., \cite{MCC,Barbarossa1}). In particular, in this work, we focus on the class of ``map-reduce"-type call graphs, illustrated in Fig. \ref{Fig1}, which is characterized by input (``map") task, to be run at the mobile; processing tasks $\mathrm{T}_i$, for $i=1,...N$, which may be offloaded; and an output task (``reduce") to be run at the mobile. As seen in Fig. 1, the directed edge between input task and task $\mathrm{T}_{i}$ is labeled by the size $b_{i}^{I}$ in bits of the data needed for task $\mathrm{T}_{i}$ to run; while the directed edge between each task $\mathrm{T}_{i}$ and the output task is labeled by the number $b_{i}^{O}$ of bits produced by the task. Furthermore, each task $\mathrm{T}_{i}$ is labeled by the number of CPU cycles $v_{i}$ required by its execution. Note that, if task $\mathrm{T}_{i}$ is offloaded, $b_{i}^{I}$ bits need to be transmitted in the uplink and $b_{i}^{O}$ bits should be received in the downlink direction.

\begin{figure}[t]
\centering \includegraphics[width=0.4\textwidth]{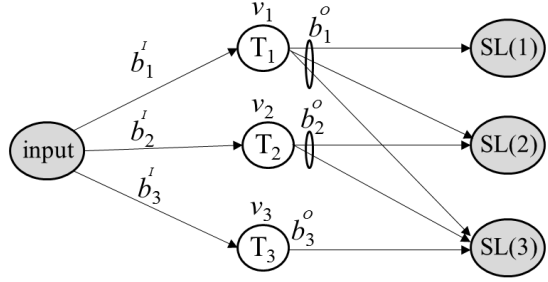} \protect\caption{Example of a compound call hypergraph.}

\label{fig:system}

\end{figure}
\textbf{Reliable Service Composition:} According to the principle of reliable service composition \cite{Popovski} (see also \cite{Dohler}), the output task can provide services corresponding to different accuracy or quality of experience levels depending on the number of tasks $\mathrm{T}_{i}$ from which it receives data. In particular, it is assumed that the tasks are ordered so that receiving from $\mathrm{T}_{1}$ allows to obtain a minimal acceptable performance, which is referred to as Service Level 1 (SL(1)); processing the inputs from $\mathrm{T}_{1}$ and $\mathrm{T}_{2}$ yields an enhanced performance, denoted as SL(2); and so on for every subset $\{\mathrm{T}_{1},...,\mathrm{T}_{i}\}$ for $i=1,...,N$, which yields a service level SL($i$), with full quality obtained when the outputs of all tasks $\{\mathrm{T}_{1},...,\mathrm{T}_{N}\}$
are available at the output task. Note that extensions in which more general nested subsets correspond to different quality of experience metrics could be accommodated in the framework.

Reliable service composition requires that the $i$th service level (SL($i$)) be obtained with probability $r_{i}$, with $r_{1}\geq r_{2}\geq...\geq r_{N}$. For example, in order to ensure ultra-reliability, SL(1) may be provided with reliability $r_{1}=99\%$, while a lower reliability may be sufficient for higher service levels. We define also the parameters $\tilde{r}_{i}=r_{i}/r_{i-1}$, with $\tilde{r}_{1}=r_{1}$, where $\tilde{r}_{i}$ measures the probability that SL($i$) is realized given that SL($i-1)$ is also attained. This follows from the definition of SLs, which imply that SL($i$) can only be realized if SL($i-1$) is.

\textbf{Offloading:} The offloading decisions are described by binary variables $I_i$. Specifically, the $i$th task can be either offloaded, which is indicated by setting $I_{i}=1$, or computed locally, indicated as $I_{i}=0$. The set of offloaded tasks is represented by $\mathcal{T}$, i.e., $\mathcal{T}=\{i\in\{1,...,N\}: I_i=1\}$. If task $\mathrm{T}_{i}$ is offloaded, a transmission power $P_{i}^{I}$ is allocated to send $b_{i}^{I}$
bits in the uplink, while $P_{i}^{O}$ is the allocated transmission power to send $b_{i}^{O}$ bits in the downlink direction. Rayleigh fading is used to model the communication channel between user and BS, with a diversity order of $d$ in both uplink and downlink. For the sake of simplicity and concreteness, selection diversity is utilized to exploit the diversity. It is assumed that mobile has no knowledge about the channel while the BS has full knowledge. Channels in the uplink and downlink direction are independent of each other. Furthermore, we let $f^{M}$ and $f^{C}$ be mobile and cloud computing frequencies,  respectively, in CPU cycles per second. We also denote as $P_M$ the power needed to compute locally at the mobile device. Finally, the application latency constraint states that maximum allowed delay, including the time need for communication and computing, is $L_{max}$ second.

\textbf{Compound Hypergraph:} To simplify the interpretation of the reliable service composition requirements, we now introduce an alternative graphical representation that we refer to as compound hypergraph. While this is not necessary for what follows, we believe it to be a useful way to visualize the reliability requirements. To elaborate, given a call graph as in Fig. 1, a compound call hypergraph can be constructed  in order to represent the reliable service composition requirements as follows:

\begin{itemize}
\item Set input node and edges between input task and tasks $\mathrm{T}_{i}$
as in original call graph;
\item Replicate the output task $N$ times, the first corresponding to SL(1),
the second to SL(2) and so on. We refer to each output node by the
corresponding service level;
\item Connect task $\mathrm{T}_{i}$ to the output nodes SL($j$) with $j=i,...,N$
via a directed hyperedge with head in $\mathrm{T}_{i}$ and tail given
by the set \{SL($i$),SL($i+1$),...,SL($N$)\}. The hyperedge is
labeled by the size of the output of task $\mathrm{T}_{i}$, namely
$b_{i}^{O}$.
\end{itemize}

Fig. 2 is an example of a compound call hypergraph. The hypergraph formalism is useful to capture the fact that, if task $\mathrm{T}_{i}$ is offloaded, $b_{i}^{O}$ bits need to be received for all the connected SL output tasks.

\textbf{Time-division vs. superposition coding} Two transmission modes are considered for offloading, namely:

\begin{itemize}
\item Time-division (TD) transmission: Input bits $b_{i}^{I}$ for $i\in\mathcal{T}$
on the uplink and output bits $b_{i}^{O}$ for $i\in\mathcal{T}$
on the downlink are transmitted in separate time slots, each of duration length $L^{I}_i$ and $L^{O}_i$, respectively. More in detail, in a time slot of duration $L^I$, the bits for all the offloaded tasks are encoded into different codewords of the same length that are summed, i.e., superimposed, for transmission in the uplink. The same is done for the downlink in a time-slot of duration $L^O$.
\item Superposition coding (SC): Input bits $b_{i}^{I}$ for $i\in\mathcal{T}$ on the uplink and output bits $b_{i}^{O}$ for $i\in\mathcal{T}$ on the downlink are transmitted using superposition coding in two separate time-slots of length $L^I$ and $L^O$, respectively. The BS in the uplink and the mobile in the downlink decode in lexicographical order starting from the bits corresponding to the lower index $i$ by means successive interference cancellation.
\end{itemize}

\subsection{Design Problem Formulation}

We focus on the problem of minimizing the energy consumption at the mobile subject to the mentioned maximum latency constraint and reliability constraints, as well as power constraints at the base station. The resulting optimization problem is stated as:
\begin{equation}
\begin{array}{llll}
\textrm{\ensuremath{\ensuremath{\textrm{minimize}}}} & \sum_{i=1}^{N} \left ( I_{i}P_{i}^{I}L^{I}_{i}+\frac{(1-I_i)v_i}{f^M}P_M \right )\\
\textrm{subject}\textrm{ to} & \sum_{i=1}^{N} \left (  I_{i}(L^{I}_{i}+L^{O}_{i})+\frac{I_{i}v_{i}}{f^C} + \frac{(1-I_i)v_i}{f^M} \right ) \leq L_{\textrm{max}}  \\
\textrm{} & \rho_{i}^{I}(\mathbf{P}^{I},L^{I}_{i},\mathbf{I}) \geq \sqrt{\tilde{r}_{i}}~~~~~~~\textrm{for } i\in \mathcal{T}\\&
\rho_{i}^{O}(\mathbf{P}^{O},L^{O}_{i},\mathbf{I}) \geq \sqrt{\tilde{r}_{i}} ~~~~~\textrm{for } i\in \mathcal{T}\\&
P_{i}^{O} \leq P_{max}^{DL} ~~~~~~~~~~~~~~~~\textrm{for } i\in \mathcal{T}
\\&
P_{i}^{I} \geq 0, P_{i}^{O} \geq 0, L_{i}^{I} \geq 0, L_{i}^{O} \geq 0 \\
& I_{i}\in\{0,1\}  \\
\textrm{variables}& \{I_{i},P_{i}^{I},P_{i}^{O},L^{I}_{i},L^{O}_{i} \}
\end{array}
\end{equation}
where $\mathbf{P}^{I}=(P_1^I,...,P_N^I)$ and $\mathbf{P}^{O}=(P_1^O,...,P_N^O)$ are the vectors of transmission powers in uplink and downlink directions, respectively; $\mathbf{I}=(I_1,...,I_N)$ is the vector collecting all the offloading decisions; $L_i^I$ and $L_i^O$ are the uplink and downlink transmission times, respectively, as introduced above. Note that problem (1) applies to both TD and SC transmissions, with the only caveat that, with SC, we have the additional constraint that $L_i^I=L^I$ and $L_i^O=L^O$ for all $i=1,...,N$. The functions $\rho_{i}^{O}(\mathbf{P}^{O},L^{O}_{i},\mathbf{I})$ and $\rho_{i}^{I}(\mathbf{P}^{I},L^{I}_{i},\mathbf{I})$ represent the probabilities of success for the transmissions in the uplink and in the downlink, respectively, for the offloading of task $T_i$. These functions depend on whether the transmission takes place via TD or SC, as further discussed below.

The objective function in (1) is the sum of transmission energy at the user, which accounts for the offloaded tasks, and of the local computing energy, for tasks that are run locally. In a similar manner, the first constraint accounts for the latency of both transmission and computing. The following reliability constraints in (1) are justified by the fact that the reliability of SL($i$), conditioned on SL($i-1$), is given by the product of the probabilities of success for uplink and downlink transmissions. This is because task $\mathrm{T}_i\in\mathcal{T}$ is successfully offloaded as long as both uplink and downlink transmissions are successful. The problem formulation in (1) is obtained by imposing equal reliability requirements on uplink and downlink. A problem formulation with a more general balancing could be easily defined, but is not further considered here. Finally, the fourth constraint imposes a power limit on the transmission of the BS, due to the power-limited, rather than energy-limited, nature of BS transmission.

Problem (1) is a mixed integer program. To solve this problem, we perform an exhaustive search over the binary variable $I_i$ and adopt the successive convex approximation method of \cite{Scutari} to optimize over the remaining variables namely $\{P_{i}^{I},P_{i}^{O},L^{I}_{i},L^{O}_{i} \}$ for fixed offloading variables. This method is invoked since, as further detailed below, for fixed offloading variables, problem (1) is not convex. For instance, the objective function of problem (1) is a non-convex bilinear function in the optimization variables $(P_{i}^{I},L^{I}_{i})$.

 We now specialize problem (1) to TD and SC transmission.

\subsubsection{Time Division Transmission}
For TD transmission, using outage capacity arguments, the probability of a successful transmission for the uplink can be written as (see Appendix):

\begin{equation}
\rho_{i}^{I}(P^{I}_i,L^{I}_{i},I_i)\hspace{-1mm}=\hspace{-1mm}  \left(\hspace{-1mm} 1-\hspace{-1mm}\left(\hspace{-1mm}1-\exp \Bigg(-\frac{2^{\frac{b_{i}^{I}}{L^{I}_{i}W^{I}}}-1}{\gamma^I P_{i}^{I}}\Bigg)\hspace{-1mm} \right)^d \right) \hspace{-1mm}
\end{equation}

\noindent and analogously for the downlink by substituting the superscript "O" for "I". In (2), $\gamma^I$ stands for average signal-to-noise ratio (SNR) of the uplink channel for a unitary transmit power, i.e., for $P_{i}^{I}=1$. We define the downlink average SNR $\gamma^O$ in a similar way. The original problem (1) can be seen to be non-convex due to the bilinearity of the objective. Using the successive convex approximation method in \cite{Scutari}, the original problem (1) is solved as outlined in Table I by means of an iterative procedure in which the current iterate is denoted as $s^{t}=\{p_{i}^{I},l_{i}^{O},l^{I}_{i},l^{O}_{i} \}$, where $t$ is the iteration index. At each iteration, the following strictly convex problem is solved:

\begin{equation}
\begin{array}{llll}
\textrm{\ensuremath{\ensuremath{\textrm{minimize}}}} & \sum_{i=1}^{N} \Bigg ( I_i\Big ( p_i^I(L_i^I-l_i^I)+\frac{\tau_i^I}{2}\|L_i^I-l_i^I\|^2 \\&  \hspace{-5mm} +l_i^I(P_i^I-p_i^I) +\frac{\tau_i^P}{2}\|P_i^I-p_i^I\|^2\Big )+\frac{(1-I_i)v_i}{f^M}P_M \hspace{-1mm}\Bigg )\\
\textrm{subject}\textrm{ to} & \hspace{-2mm}\sum_{i=1}^{N} \left (  I_{i}(L^{I}_{i}+L^{O}_{i})+\frac{I_{i}v_{i}}{f^C} + \frac{(1-I_i)v_i}{f^M} \right ) \leq L_{\textrm{max}}  \\&
\hspace{-2mm}\frac{2^{\frac{b_{i}^{I}}{L^{I}_{i}W^{I}}}-1}{\gamma^I P_{i}^{I}} + \ln(1-(1-\sqrt{\tilde{r}_{i}})^{\frac{1}{d}}) \leq 0 ~\textrm{for } i\in \mathcal{T}\\&
\hspace{-2mm}\frac{2^{\frac{b_{i}^{O}}{L^{O}_{i}W^{O}}}-1}{\gamma^O P_{i}^{O}} + \ln(1-(1-\sqrt{\tilde{r}_{i}})^{\frac{1}{d}}) \leq 0 ~\textrm{for } i\in \mathcal{T} \\&
\hspace{-2mm}P_{i}^{O} \leq P_{max}^{DL}    ~~~~~~~~~~~~~~~~~\textrm{for } i\in \mathcal{T}\\&
\hspace{-2mm}P_{i}^{I} \geq 0, P_{i}^{O} \geq 0, L_{i}^{I} \geq 0, L_{i}^{O} \geq 0 \\
\textrm{variables}& \hspace{-2mm} \{P_{i}^{I},P_{i}^{O},L^{I}_{i},L^{O}_{i} \} \hspace{-2mm}
\end{array}
\end{equation}
\noindent Note that all the constraints in the problem above are convex. Also, the second and third constraints are obtained from simple algebraic manipulations from the corresponding constraints in (1). In Table I, the step sizes are updated as $\lambda^{t+1}=\lambda^{t}(1-\epsilon\lambda^{t})$ for $t \geq 0$ with $\lambda^0 \in (0,1]$ and $\epsilon^0 \in (0,1)$. The algorithm in Table I is repeated until convergence for every value of $I_{i}$. The minimum value of the objective over all possible choices of $I_{i}$ is taken as the final solution.

\begin{table}[b]
\centering
\caption{Successive convex approximation algorithm for TD}
\label{Table:TD}
\begin{small}
\begin{tabular}{|l|}
  \hline
  Initialization: Set $t=0$, $s^0$=$\{p_{i}^I,p_{i}^O,l_{i}^I,l_{o}^I\}$ feasible \\
  ~~~~~~~~~~~~~~~~$\lambda^0 \in (0,1]$, $\epsilon^0 \in (0,1)$ \\
  \hline
  Step 1) If $s^t$ satisfies a termination criterion: STOP \\
  Step 2) Compute $\hat{s}(s^t)$ as the solution of (3).  \\
  Step 3) Set $s^{t+1} = s^{t} +\lambda^t(\hat{s}(s^t)-s^{t})$. \\
  (S.4) $t \leftarrow t + 1$ and go to step 1.\\
  \hline
\end{tabular}
\end{small}
\end{table}

\subsubsection{Superposition Coding Transmission}

With SC, the probability of success for uplink transmission can be written as (see Appendix):

\begin{equation}
\begin{array}{llll}
\hspace{-1mm}\rho_{i}^{I}(P^{I}_i,L^{I},I_i)\hspace{-1mm}=\hspace{-1mm} 1-\hspace{-1mm}\left(\hspace{-1mm}1\hspace{-1mm}-\exp\hspace{-1mm}\left(\hspace{-1mm}\frac{-\Big(2^{\frac{b_{i}^{I}}{L^{I}W^{I}}}-1\Big)}{\gamma^I P_{i}^{I}-\hspace{-1mm}\left(\hspace{-1mm}2^{\frac{b_{i}^{I}}{L^{I}W^{I}}}-1\hspace{-1mm}\right)\hspace{-1mm}\sum_{j=i+1}^{N}\gamma^I P_{j}^{I}\hspace{-1mm}}\hspace{-1mm}\right)\hspace{-2mm} \right)^d \hspace{-3mm}
\end{array}
\end{equation}

\noindent and analogously for the downlink. Note that here the transmission periods $L^I$ and $L^O$ do not depend on the task $i$ as explained above. As for TD, the reliability constraint can be expressed as
\begin{equation}
\begin{array}{llll}
 \frac{2^{\frac{b_{i}^{I}}{L^{I}W^{I}}}-1}{\gamma^I P_{i}^{I}}\hspace{-1mm} -\hspace{-1mm}\frac{1}{\hspace{-3mm}\gamma^I \hspace{-1mm}\sum_{j=i+1}^{N}P_j^I-\Big(\ln(1-(1-\sqrt{\tilde{r}_{i}})^{\frac{1}{d}})\Big)^{-1}} \leq 0
 \end{array}
\end{equation}
\noindent for the uplink and analogously for the downlink. Unlike TD, these constraints are non-convex. However, they can be written as the difference of two convex functions, which may be dealt with as explained in \cite{Scutari} in the successive convex approximation method by linearizing the negative term. This yields the approximate reliability function:

\begin{equation}
\begin{array}{llll}
\frac{2^{\frac{b_{i}^{I}}{L^{I}W^{I}}}-1}{\gamma^I P_{i}^{I}} -\frac{1}{\gamma^I \sum_{j=i+1}^{N}p_j^I-\Big(\ln(1-(1-\sqrt{\tilde{r}_{i}})^{\frac{1}{d}})\Big)^{-1}} \\  +\frac{\gamma^I\Big(\sum_{j=i+1}^{N}( P_{j}^{I}-p_{j}^{I})\Big)}{\Big(\sum_{j=i+1}^{N}\gamma^I p_j^I-\Big(\ln(1-(1-\sqrt{\tilde{r}_{i}})^{\frac{1}{d}})\Big)^{-1}\Big)^2} \hspace{-1mm} \leq  0
\end{array}
\end{equation}
where $\{p_{i}^{I}\}$ represents the previous iterate. Following Table I, the problem to be solved at each iteration is then (3), with $L_{i}^I=L^I$ and $L_{i}^O=L^O$ as well as with the constraint above, and the corresponding downlink constraint, in lieu of the third and fourth constraints in (3).

\section{Numerical Results}
In this section, we provide some numerical examples based on the analysis developed in the previous sections. We set $P_{M}= 0.4$ Watts; $f_M = 10^9$ CPU cycles/s (e.g., Apple iPhone 6 processor has maximum clock rate of $1.4$ GHz); $f_C = 10^{10}$ CPU cycles/s (e.g., AMD FX-9590 has a clock rate of 5 Ghz); bandwidth $W^{I}=1 MHz$ and $W^{O}=1 MHz$;  and SNR levels $\gamma^I=\gamma^O=0$ dB.

We start by considering just the basic service level, namely SL(1), in order to simplify the interpretation of the results and to gain insight into the role of the diversity level $d$ on the offloading decisions. The reliability of SL(1) is set to $r_1=0.99$. Note that, given the presence of only one offloadable task, TD and SC yield the same performance. Fig. 3 presents the mobile energy versus latency constraint $L_{max}$ for different diversity orders. If the tolerable latency is low, then it is necessary to offload the task to the cloud since local computing here takes around $1.7$ seconds. For diversity $d=1$, the energy required to offload is outside the range shown in the Fig. 1. An increase of diversity order provides more reliable communication between mobile and cloud, and therefore the offload of the task can be performed with a lower energy expenditure. In particular, if $d = 3$, then it is optimal to offload the task even for latencies larger than 1.7 seconds. We emphasize that the discontinuity in the curves is due to changes in the optimal offloading decisions.

\vspace{4mm}
\begin{figure}[t]
\centering \includegraphics[width=0.45\textwidth]{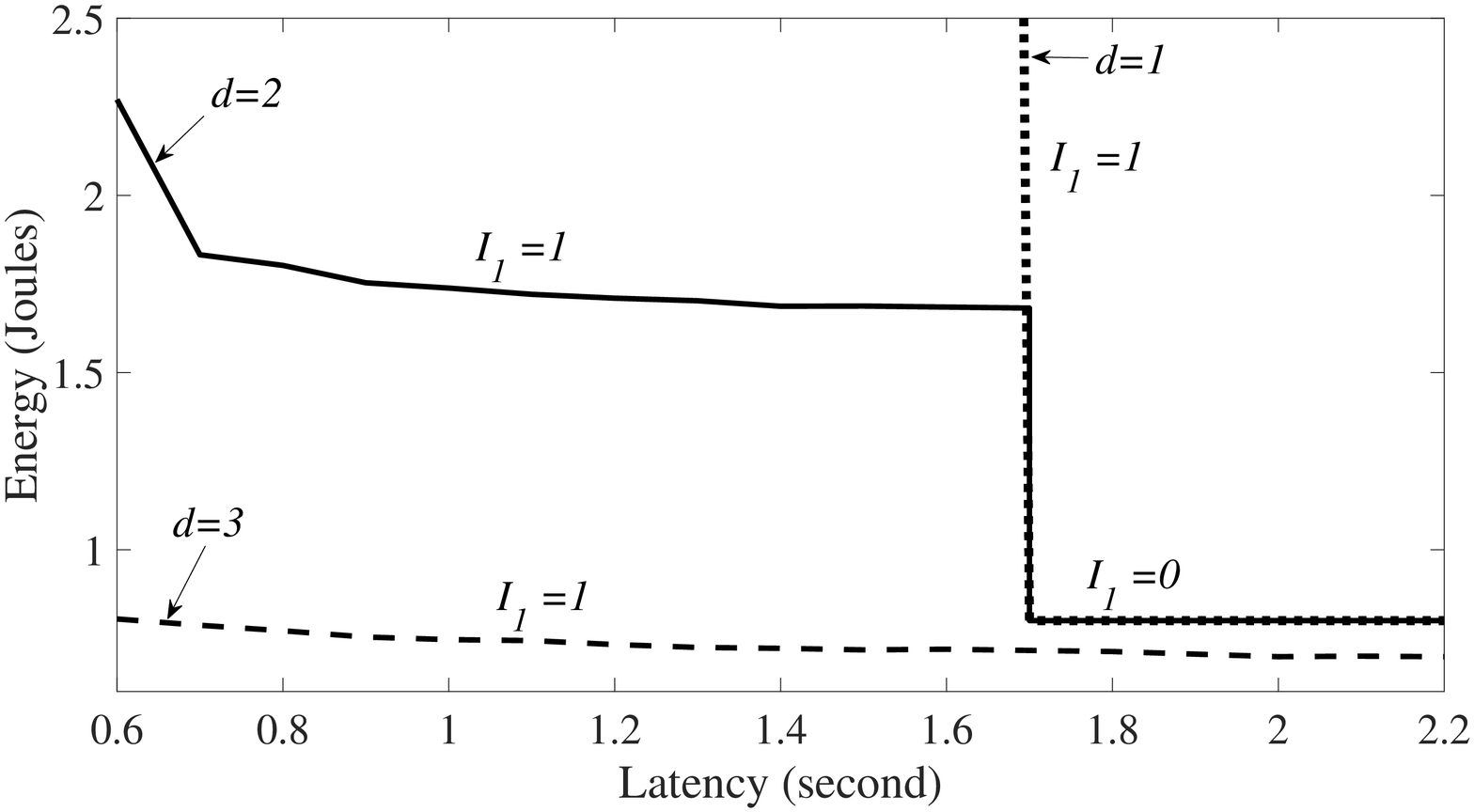} \caption{Mobile energy expenditure versus latency constraint for a single service level ($v=2 \times 10^{9}$ CPU cycles and $b_1^I=b_1^O=1.4 \times 10^{5}$ bits).}

\label{fig:system}

\end{figure}

We now consider reliable service composition with two service levels, namely $N = 2$, accounting for both TD and SC transmission modes. The reliability for the second level is set to $\tilde{r}_2=0.9$ and the first is still $r_1=0.99$. The corresponding mobile energy versus latency trade-offs are shown in Fig. 4 and Fig. 5, respectively. Note that here computing both tasks locally requires a latency of approximately $3.6$ seconds. Considering first TD transmission, we observe from Fig. 4 that achieving low latency requires tolerating a high energy cost by offloading both tasks. When increasing the latency, the mobile has incentive to first offload the task with higher computation cost, here the first task, while the second task is run locally due to the lower energy consumption. For $d = 2$ and higher latencies, when the first task can be run locally, it becomes optimal to offload only the second task; while, when the latency is large enough, both tasks should be run locally. With a larger latency, instead, the solution ($I_1 = 1, I_2 = 0$) turns out to be optimal over a larger range of latencies.

Comparing TD with SC, by observing Fig. 5, we note that SC enables a drastic energy reduction for offloading and hence makes the decision to offload both tasks optimal for all latencies up to $3.6$ seconds when $d=2$, and for the entire range of considered latencies when $d=3$.

\vspace{4mm}
\begin{figure}[t]
\centering \includegraphics[width=0.5\textwidth]{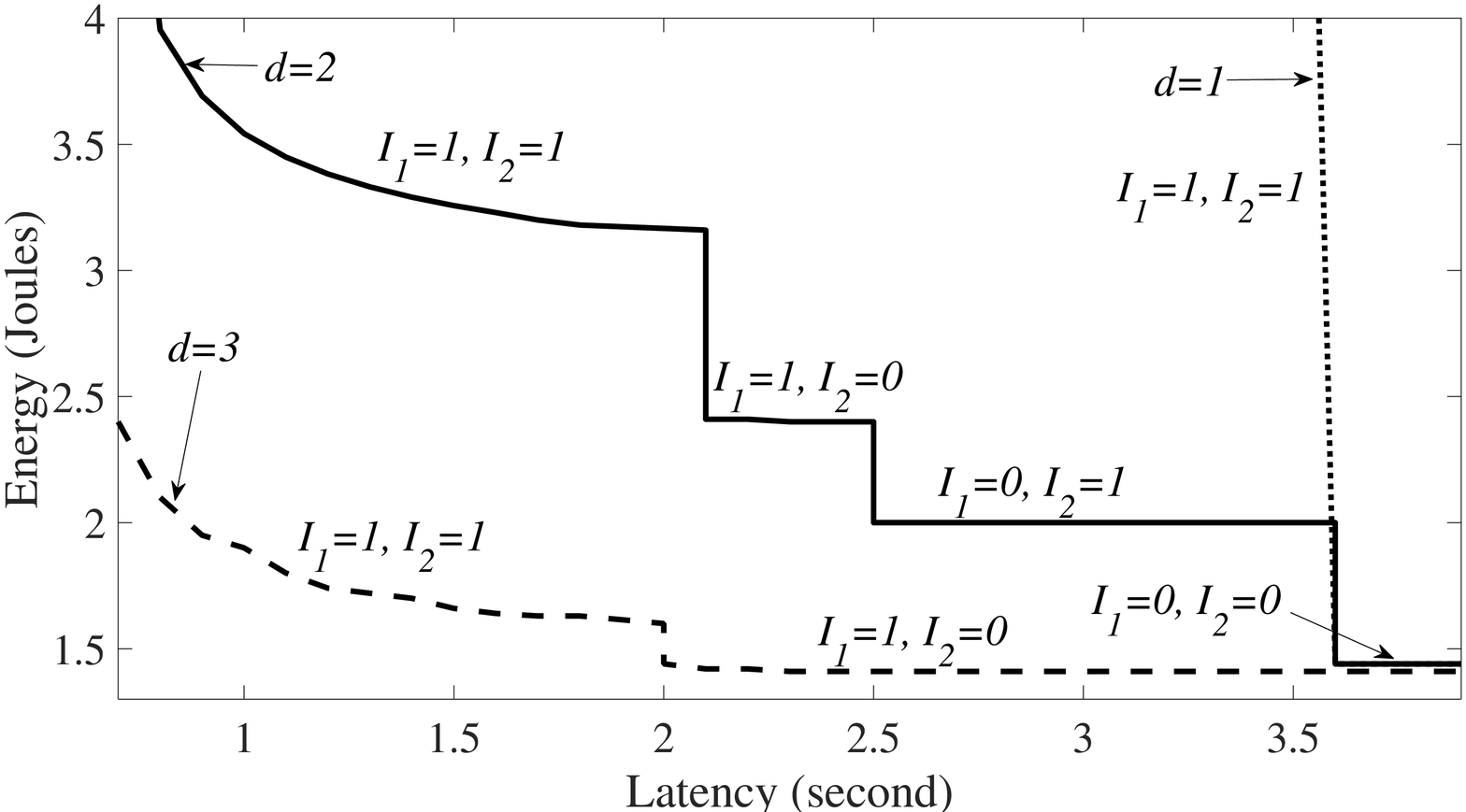} \caption{Mobile energy expenditure versus latency constraint for a single service level ($v_1=2 \times 10^{9}$ CPU cycles, $v_2=1.6 \times 10^{9}$ CPU cycles, $b_1^I=b_1^O=1.4 \times 10^{5}$ bits and $b_2^I=b_2^O=2.8 \times 10^{5}$ bits).}

\label{fig:system}

\end{figure}

\begin{figure}[b]
\centering \includegraphics[width=0.45\textwidth]{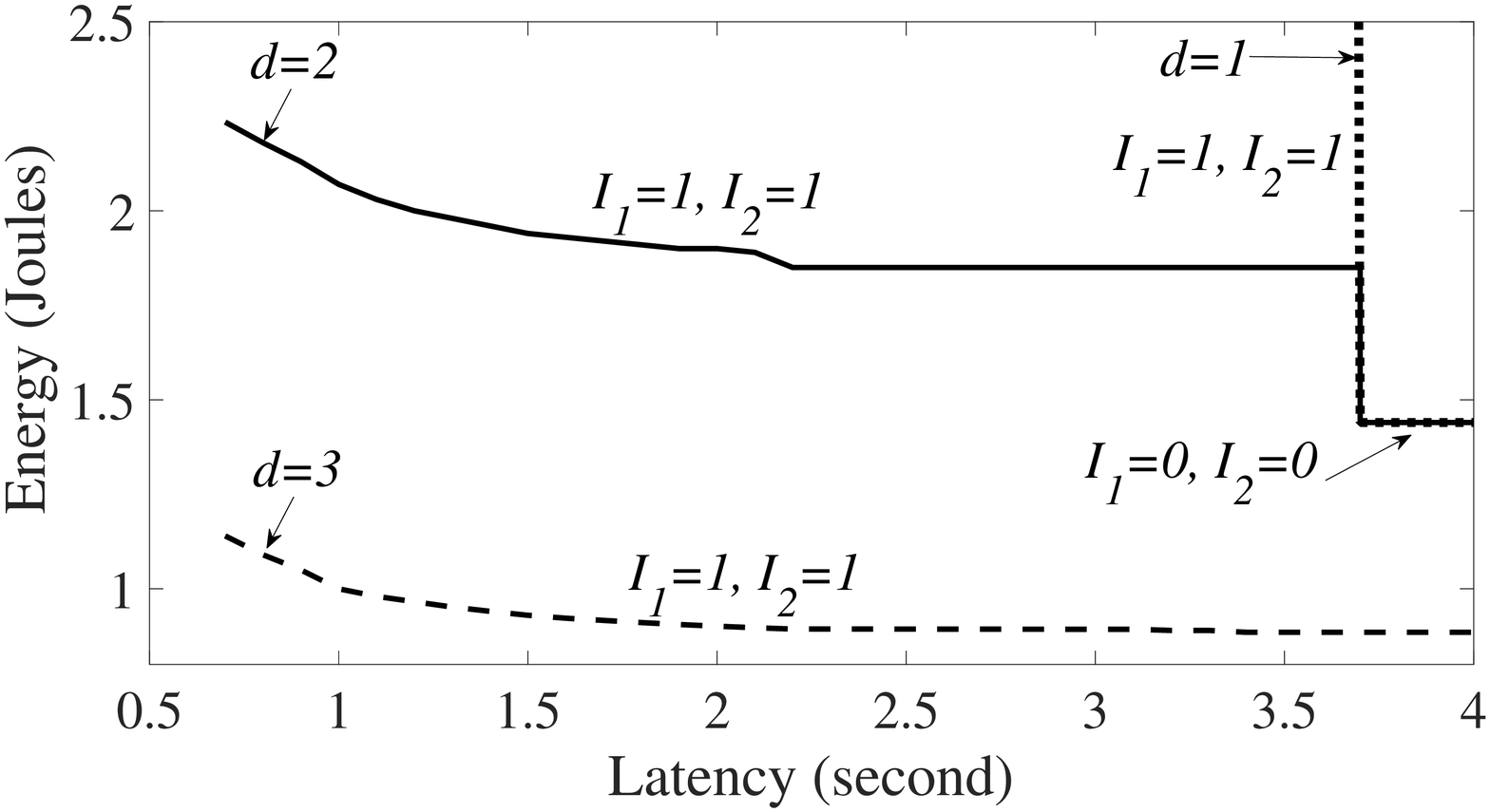} \caption{Mobile energy expenditure versus latency constraint for a single service level ($v_1=2 \times 10^{9}$ CPU cycles, $v_2=1.6 \times 10^{9}$ CPU cycles, $b_1^I=b_1^O=1.4 \times 10^{5}$ bits and $b_2^I=b_2^O=2.8 \times 10^{5}$ bits).}

\label{fig:system}

\end{figure}

\section{Concluding Remarks}
In this paper, the mobile energy versus latency tradeoff was explored for mobile cloud computing applications over fading channels by accounting for the principle of reliable service level composition at the application layer. The aim of this approach is providing layered, or composable, services at differentiated reliability levels. Inter-layer optimization problems, encompassing offloading decisions and communication resources, were formulated and addressed by means of successive convex approximation methods. The numerical results demonstrated the energy savings that may be obtained by a joint allocation of computing and communication resources, as well as the advantages of superposition coding at the physical layer and the impact of channel conditions on the offloading decisions.


\section*{Appendix: Calculation of the Probabilities of Success}
For TD transmission, assuming diversity $d=1$ and Rayleigh fading channel gain $G^I$, the probability of success is the complement of the outage probability, namely
\begin{equation}
\begin{array}{ll}
 \mbox{Pr\ensuremath{\left[L^{I}_{i}W^{I}\log(1+G^{I}P_{i}^{I}\gamma^I)\geq b_{i}^{I}\right]}} = & \\
 \textrm{Pr}\ensuremath{\left[G^{I}\geq\frac{2^{\frac{b_{i}^{I}}{L^{I}_{i}W^{I}}}-1}{\gamma^IP_{i}^{I}}\right]}   =\exp\left(-\frac{2^{\frac{b_{i}^{I}}{L^{I}_{i}W^{I}}}-1}{\gamma^IP_{i}^{I}}\right).
\end{array}
\end{equation}

Generalizing, with a diversity order $d \geq 1$ and selection diversity, we obtain

\begin{equation}
\begin{array}{ll}
\rho_{i}^{I}(P_{i}^{I},L^{I}_{i},I_i) =
 1-\left(1-\exp \Bigg(-\frac{2^{\frac{b_{i}^{I}}{L^{I}_{i}W^{I}}}-1}{\gamma^IP_{i}^{I}}\Bigg) \right)^d,
\end{array}
\end{equation}
and similar calculations apply for $\rho_{i}^{O}(P_i^{O},L^{O}_{i},I_i)$.

For SC transmission, assuming for $d=1$, following similar arguments, we have

\begin{equation}
\begin{array}{ll}
\rho_{i}^{I}(\mathbf{P}^{I},L^{I},\mathbf{I}) \nonumber \\
= \mbox{Pr\ensuremath{\left[L^{I}W^{I}\log_{2}\left(1+\frac{G^{I}P_{i}^{I}\gamma^I}{1+\gamma^IG^{I}\sum_{i=i+1}^{N}P_{i}^{I}}\right)\geq b_{i}^{I}\right]}}\nonumber \\
=  \textrm{Pr}\ensuremath{\left[G^{I}P_{i}\gamma^I\geq\left(2^{\frac{b_{i}^{I}}{L^{I}W^{I}}}-1\right)\left(1+\gamma^IG^{I}\sum_{i=i+1}^{N}P_{i}^{I}\right)\right]}\nonumber \\
= \textrm{Pr}\ensuremath{\left[G^{I}\geq\frac{2^{\frac{b_{i}^{I}}{L^{I}W^{I}}}-1}{\gamma^IP_{i}-\left(2^{\frac{b_{i}^{I}}{L^{I}W^{I}}}-1\right)\gamma^I\sum_{i=i+1}^{N}P_{i}^{I}}\right]}\nonumber \\
= \exp\left(-\frac{2^{\frac{b_{i}^{I}}{L^{I}_{i}W^{I}}}-1}{\gamma^IP_{i}-\left(2^{\frac{b_{i}^{I}}{L^{I}W^{I}}}-1\right)\gamma^I\sum_{i=i+1}^{N}P_{i}^{I}}\right),
\end{array}
\end{equation}
where all layers beyond $i$ are treated as noise in the decoding. With selection diversity, we obtain the reliability function stated in the text.
\balance

\bibliographystyle{IEEEtran}
\bibliography{IEEEabrv,IEEEexample}
%\begin{thebibliography}{1}


%\end{thebibliography}

\end{document}